\documentclass[%
 reprint,
 nofootinbib,
 amsmath,amssymb,
 aps,
]{revtex4-1}

\usepackage{graphicx}
\usepackage{dcolumn}
\usepackage{bm}
\usepackage{xcolor}
\usepackage[colorlinks=true,linkcolor={blue!50!black},citecolor={blue!50!black},urlcolor={blue!80!black}]{hyperref}
\usepackage{mathtools} 

\DeclarePairedDelimiter{\ceil}{\lceil}{\rceil}
\DeclarePairedDelimiter{\floor}{\lfloor}{\rfloor}

\begin{document}

\title{Fermions in Rotating Reference Frames}

\author{Adrian Manning}
\email{a.manning@physics.usyd.edu.au}
\affiliation{ARC Centre of Excellence for Particle Physics at the Terascale, School of Physics, The University of Sydney, Sydney, NSW 2006, Australia}

\date{December 2015}

\begin{abstract}
Current quantisations of fermions in cylindrical coordinates are shown to be inadequate in calculating some single-particle expectation values. This paper develops an alternate quantisation, applicable to one-particle states, which is generalised to rotating frames in cylindrical coordinates. Using this quantization, an explicit calculation of the velocity of a flat space free fermion as observed by a rotating observer is presented. This calculation demonstrates the validity of this quantisation and the cancellation of non-inertial rotational effects in the velocity, naively expected from the mixing of particle and anti-particle field operators.
\end{abstract}

\maketitle

\section{Introduction}  

The quantisation of fields in non-Minkowskian metrics can give rise to observable effects, not otherwise present in the flat-space formalism. The most notable examples are Hawking's black-hole evaporation \cite{Hawking1975} and Fulling-Davies-Unruh's \cite{Fulling1973}\cite{Davies1975}\cite{Unruh1976} accelerating thermal bath. These semi-classical gravitational and inertial effects occur due to the differing operators that arise when quantizing fields in inequivalent reference frames. Different quantum operators necessarily imply different vacuum states between the two quantisations of the field. Linking the two quantisations via Bogoliubov transformations \cite{Birrell1984} can give extra contributions to standard flat-space observer expectation values when dealing with fields in non-Minkowskian metrics. Specific examples of these effects can be found in the references  \cite{Kobakhidze2015a}\cite{Kobakhidze2015}\cite{Giovannini2000}\cite{Girdhar2013} and in the reviews  \cite{Birrell1984}\cite{Jacobson:2003vx}\cite{Parker2009}.

Due to these effects, it is important to be able to quantise fermionic matter fields in common metrics in order to estimate any deviation to expectation values in standard Minkowskian quantum field theory (QFT) calculations. This paper focuses on the case of the non-inertial metric associated with rotation, which can have useful application to calculations involving rotating particles such as in storage rings, synchrotron's and various cosmological and astrophysical scenarios (e.g. \cite{Dvornikov2014}).

Some specific quantisations for spinor fields in rotating frames have been developed (see for example \cite{Iyer1982}, \cite{Vilenkin1980}, \cite{Ambrus2014}), however these follow quantisations completed in flat-space using cylindrical coordinates (\cite{Audretsch1996}, \cite{Skarzhinsky1997}, \cite{Balantekin}) where the field quanta represent standing waves composed of radially in-going and out-going components. These quantisations are inadequate to describe free one-particle states in rotating and flat space scenarios as the quanta are not representative of free single state Minkowskian particles with a definite trajectory. A clear example of the inadequacy of these quantisations is given through the calculation of the radial current density of a  one particle state in these quantisations. This calculation is shown to vanish for all quantum numbers of the one particle state (demonstrated in Section \ref{sec:flatspacecylin}). This result is an artefact of choosing the boundary condition such that the field vanishes at the origin. However, one should bear in mind that the origin in cylindrical coordinates is associated with a coordinate divergence and has no relevance to physics.

In order to adequately calculate one-particle expectation values, a new quantisation is required. I propose an alternate quantisation where the quanta represents a one particle state with definite radial momenta which can be intuitively matched to a single particle in flat Minkowskian space. This quantization applies equally well to flat or rotating spaces in cylindrical coordinates. 

In the generalised rotating case, the Bogoliubov coefficients are calculated between the rotating and flat space quantisations allowing arbitrary expectation values to be calculated for the alternate quantisation in the rotating frame. Of particular interest, is the presence of a non-zero $\beta$ coefficient. This suggests a mixing between the flat space particle and anti-particle states in the rotating frame. Naively, this would imply an oscillatory motion (Zitterbewegung \cite{Kobakhidze2015a}) when estimating the velocity of a Minkowskian one particle state in the rotating frame. Through explicit calculation of the velocity expectation value using the new quantisation, we show that the non-inertial effects of rotation cancel out, leading to no observable Zitterbewegung effect due to rotation.  

\section{Standard Quantisation of Fermions in a Rotating Frame}
The line element used to specify the rotating frame, which is rotating at a constant angular velocity, $\omega$, relative to an inertial observer is (in cylindrical coordinates)
\begin{equation}
	ds^2 = (1 - \omega^2 r^2) dt^2 -dr^2 -r^2 d\theta^2 -2\omega r^2 d\theta dt - dz^2 \;.
	\label{eq:metric}
\end{equation}
The Dirac equation in a general spacetime can be written as 
\begin{equation}
	i e_a^\mu \gamma^a \nabla_\mu \psi - m\psi = 0 \;.
\end{equation}
Here Latin indices and Greek indices represent flat and curved space Lorrentz indices respectively and $m$ is the mass of the fermion, $\psi$. The tetrad $e_\mu^a$ is introduced to allow the Dirac gamma matrices to be written in their flat space representation 
\begin{equation}
	\left\{ \gamma^a , \gamma^b \right\} = 2 \eta^{ab} \;.
\end{equation}
The spinor affine connection $\Gamma_\mu$ is contained within the covariant derivative as follows
\begin{equation}
	\nabla_\mu \psi = \partial_\mu \psi + \Gamma_\mu \psi \;,
\end{equation}
and is determined by the choice of tetrads. For this metric, it is convenient to use the cylindrical tetrad gauge as in \cite{Iyer1982}.
In this gauge, the non-vanishing tetrad components are (lower indices representing rotating coordinates)
\begin{equation}
	\begin{aligned}
		e_t^0=e_r^1=e_z^3=1, \\
		e_\theta^0=\omega r \quad e_r^2 = r ,
	\end{aligned}
	\label{eq:tetrad} 
\end{equation}
which give the following non-zero components for the spinor affine connection
\begin{equation}
	\Gamma_t = \frac{\omega}{2} \gamma^1 \gamma^2, \quad \Gamma_\theta = \frac{1}{2} \gamma^1 \gamma^2  \;.
\end{equation}
Thus the Dirac equation in the rotating metric \eqref{eq:metric} using the cylindrical tetrad gauge \eqref{eq:tetrad}
can be written as
\begin{equation}
	\begin{aligned}
		&\Big[i \gamma^0 \left( \partial_t - \omega \partial_\theta \right) \\
		&+ i \gamma^1 \partial_r + i \frac{\gamma^2}{r} \left( \partial_\theta - i S_3 \right) + i \gamma^3 \partial_z - m \Big]\psi = 0 \;,
	\end{aligned}
	\label{eq:eom}
\end{equation}
where $S_3 = \frac{i}{2} \gamma^1 \gamma^2$ is the spin operator in the z-direction. It should be noted that a different gauge choice for the tetrad, such as the Cartesian gauge in \cite{Ambrus2014} will differ from the solutions presented here by a rotation in spinor space. 
In specifying the solutions to this equation, we fix the spinor structure to be eigenvectors of the following four commuting operators
\begin{equation}
	\begin{gathered}
		\hat{H} = i \partial_t, \quad \hat P_z = -i \partial_z \\
	\quad \hat J_z = -i\partial_\theta, \quad \hat S_z = -\gamma^5\gamma^3 - i \gamma^5 \frac{\partial_z}{m} \;.
\end{gathered}
	\label{eq:operators}
\end{equation}
Here $\hat J_z$ gives the total angular momentum, which has no spin dependent term (opposed to \cite{Audretsch1996}), usually associated with total angular momentum, due to the gauge choice of the tetrads. Hence the eigenvalue of this operator is set to be $l + \frac{1}{2}$ where $l$ is an integer, corresponding to a half-integer spin fermion. The $\hat S_z$ operator gives the spin projection along the z-axis and has eigenvalues 
\begin{equation}
	s_z = s \frac{\sqrt{m^2 + p_z^2 }}{m}, \quad s = \pm 1,
	\label{eq:sz}
\end{equation}
with $p_z$ being the eigenvalue of the $\hat P_z$ operator. With these operators specified, using the chiral basis for the gamma matrices and imposing the boundary condition that the field is regular at $r=0$, the solutions of \eqref{eq:eom} are

\begin{equation}
	\psi_{E,l,p_z,s}^+ = \frac{e^{-i\tilde{E}t + i(l+\frac{1}{2})\theta + ip_z z}}{4 \pi\sqrt{m }\sqrt{s_z}}\begin{pmatrix} m (ms_z+ p_z)^{-\frac{1}{2}} \phi_{E,l,p_z,s}(r) \\ (ms_z + p_z)^{\frac{1}{2}} \;\phi^*_{E,l,p_z,s}(r) \end{pmatrix} \;,
	\label{eq:sol}
\end{equation}
where the bi-spinor $\phi_{E,l,p_z,s}$ is defined as:
\begin{equation}
	\phi_{E,l,p_z,s}(r) = \begin{pmatrix}\frac{\sqrt{|E|}}{\sqrt{E}} \sqrt{E + ms_z} \; J_l (p_\perp r) \\[0.2cm] -i \frac{\sqrt{E}}{\sqrt{|E|}}\sqrt{E-ms_z}\; J_{l+1}(p_\perp r) \end{pmatrix} \;,
	\label{eq:bispin}
\end{equation}
and $p_\perp = \sqrt{E^2 - m^2 -p_z^2}$ is the perpendicular momentum, $J_l$ are Bessel functions of the first kind and $\tilde{E} = E - \omega \left(l + \frac{1}{2}  \right)$ is the eigenvalue of the Hamiltonian and hence the energy as defined in the rotating frame. This energy is shifted by $\omega (l + \frac{1}{2})$ compared to a non-rotating observer.

It should be noted that the general solution to \eqref{eq:eom} also involves Bessel functions of the second kind, which are divergent at the origin. The boundary condition imposing the field to be regular at the origin removes these divergent solutions leading to a field that is quantised as standing waves with zero radial current.

The negative energy solutions can be obtained via the charge conjugation operator
\begin{equation}
	\psi_{E,l,p_z,s}^- = C \overline{\psi}^T = i \gamma^2 \gamma^0 {\gamma^0}^T {\psi^*}^+_{E,l,p_z,s} \;.
	\label{eq:chargec}
\end{equation}
These states are normalised using the Dirac inner product
\begin{equation}
	\langle \psi, \chi \rangle = \int d^3x \sqrt{-g} \; \psi^\dagger(x) \chi(x) \;,
	\label{eq:diracprod}
\end{equation}
such that
\begin{equation}
	\begin{aligned}
		\langle \psi^\pm_{E,l,p_z,s} , \psi^\pm_{E',l',p_z',s'} \rangle &= \delta_{ss'}\; \delta_{ll'}\; \delta(p_z - p_z') \delta(E - E') \\
	\langle \psi^\pm_{E,l,p_z,s} , \psi^\mp_{E',l',p_z',s'}  \rangle &= 0 \;.
\end{aligned}
	\label{eq:norm}
\end{equation}
Using these mode functions it is possible to construct the quantum field
\begin{equation}
	\begin{aligned}
	&\hat \Psi_R(x) = \\
	&\sum_{l=-\infty}^{\infty} \int_{\tilde{E} >0, |E| \ge m}^\infty dE  \int_{-p}^p dp_z \sum_{s=\pm 1} \left[\hat a_{j} \, \psi^+_j(x) + {\hat b_j}^\dagger \, \psi^-_j(x) \right] \;.
\end{aligned}
\label{eq:standardquantisation}
\end{equation}
Here we have simplified the notation by compacting all the quantum numbers to a single index, $j = \{E,\theta_p,p_z,s\}$. We have also introduced the flat space momentum, $p = \sqrt{E^2-m^2}$. The quantization derived here, is similar to those typically used to quantise fermionic fields in a rotating frame \cite{Iyer1982}\cite{Vilenkin1980}\cite{Ambrus2014} and in cylindrical coordinates \cite{Audretsch1996}\cite{Skarzhinsky1997}. Note that through the limits placed on the energy integral, specific $l$ modes have been excluded to ensure the energy-momentum relationship is preserved. 

The energy of a rotating particle, $\tilde{E} = E - \omega(l + \frac{1}{2})$, which is always positive, allows the quantum number, $E$, to be negative for certain values of $l$. This quantum number partly defines operators $\{\hat a_j , \hat b_j\}$. In a non-rotating frame, $E$ is necessarily greater than zero. The fact that rotation allows this number to become negative means that for certain values of $l$, flat space particle operators will behave as anti-particle operators in the rotating frame. This will lead to non-zero $\beta$ Bogoliubov coefficients between rotating and flat space operators in Section \ref{sec:Bogoliubov}.

\section{Shortcomings of the standard quantisation}
\label{sec:flatspacecylin}
In this section, for clarity and without loss of generality, we consider the above quantisation in flat space cylindrical coordinates without rotation, i.e $\omega = 0$, $\tilde E \rightarrow E$. These solutions are similar to those found in the literature (e.g. \cite{Audretsch1996}\cite{Skarzhinsky1997}) for fermionic fields in cylindrical coordinates. To demonstrate the difficulty in using this quantisation for calculating expectation values of typical cartesian one-particle plane-wave states, the radial current density is calculated. Specifically, 
\begin{equation}
	\begin{aligned}
	&\langle 1_{E,l,p_z,s} |\, \hat J^1\, | 1_{E,l,p_z,s} \rangle \\
	=\,&\langle 1_{E,l,p_z,s} |\, r \, \hat \Psi^\dagger \gamma^0 \gamma^1 \hat \Psi \, | 1_{E,l,p_z,s} \rangle ~,
\end{aligned}
\end{equation}
with the $\hat \Psi$ operators being defined as in \eqref{eq:standardquantisation} with $\omega = 0$. 
Explicitly this becomes,
\begin{equation}
	\begin{aligned}
	&\langle 1_{E,l,p_z,s} |\,r \, \hat \Psi^\dagger \gamma^0 \gamma^1 \hat \Psi\, | 1_{E,l,p_z,s} \rangle \\
	&= \frac{i p_\perp}{8 \pi^2} \left[  J_{l+1} (p_\perp r) J_{l} (p_\perp r) - J_{l} (p_\perp r) J_{l+1}(p_\perp r) \right] \\
	&= 0 ~.
	\end{aligned}
	\label{eq:oldcurrent}
\end{equation}
Regardless of energy and angular momenta, all quanta in this quantisation have a vanishing radial current. The quanta in this quantisation represent cylindrical standing waves composed of radially out-going and in-going waves. This result naturally arises from the choice of boundary condition when specifying the mode solutions. If one wishes to calculate how cartesian plane-wave solutions behave in a cylindrical or rotating frame, it is convenient to use an alternate quantisation, which is formulated in the following section. 

\section{Quantisation of a single particle in a Rotating Reference Frame}

To find mode functions whose quanta can describe the typical cartesian plane-wave mode solution in the rotating frame we parameterize the cartesian momentum by an angle, $\theta_p$, in the plane of rotation as follows
\begin{equation}
	p_x = p_\perp \cos \theta_p, \quad p_y = p_\perp \sin \theta_p \;.
\end{equation}
Then noting the Jacobi-Anger expansion \cite{Gradshteyn2007}
\begin{equation}
	e^{-i(p_x x + p_y y)} = \sum_{n=-\infty}^{\infty} e^{in(\theta -(\theta_p + \frac{\pi}{2}))} J_n (p_\perp r) \;,
\end{equation}
we see that plane wave solutions can be constructed from infinite sums over the angular momenta defining the quanta in the previous quantisation \eqref{eq:standardquantisation}. We can construct new mode functions as infinite sums of angular momenta to represent a plane wave quanta. These quanta will have the angular momentum quantum number, $l$, replaced with the parameterized transverse momentum angle, $\theta_p$. To take the infinite sum of the mode functions, a redefinition of the energy is required in order to ensure the summed mode functions remain an eigenvector of the Hamiltonian. This can be done in the following way 
\begin{equation}
	\begin{aligned}
		&\psi_{\Omega,\theta_p,p_z,s}^+ =\\
		&\frac{1}{\sqrt{2 \pi^{{3}}}} \sum_{l\le\xi_-, l\ge \xi_+}^{\infty} \frac{\sqrt{|\hat E|}}{\sqrt{\hat E}}e^{-i\Omega t + i(l+\frac{1}{2})(\theta - (\theta_p + \frac{\pi}{2})) + ip_z z}\; u_{j}(r) \;.
	\end{aligned}
\end{equation}
The above mode functions, contain the spinor components
\begin{equation}
	u_j(r) = \begin{pmatrix} \frac{\sqrt{m}}{\sqrt{s_z}\sqrt{ms_z+ p_z}} \phi_{\hat E,l,p_z,s}(r) \\ \frac{\sqrt{ms_z + p_z}}{\sqrt{s_zm}} \;\phi^*_{\hat E,l,p_z,s}(r) \end{pmatrix} \;,
\end{equation}
where the bi-spinor $\phi_{\hat E,l,p_z,s}$ is defined as in \eqref{eq:bispin}. Here $\hat E = \Omega + \omega(l + \frac{1}{2})$ which now gives an $l$-dependent form for the radial momentum in \eqref{eq:bispin}, e.g $p_\perp = \sqrt{(\Omega+\omega(l + \frac{1}{2}))^2 - m^2 - p_z^2}$. Further we have introduced $\xi_\pm = \pm \frac{\sqrt{m^2 + p_z^2} -E}{\omega} - \frac{1}{2}$ which are required to restrict the $l$ summation to persevere the energy momentum relation as was done previously through the integral limits in the mode functions \eqref{eq:standardquantisation}. The negative energy solutions are again found using the charge conjugation operator in \eqref{eq:chargec}.

This new mode function, which also satisfies the equation of motion \eqref{eq:eom}, has shifted the $\omega$ dependence from the exponential phase factor containing $\tilde{E}$ in \eqref{eq:sol} to the bi-spinor components \eqref{eq:bispin} to ensure the mode function remains an eigenvector of the Hamiltonian
\begin{equation}
	\hat H \psi = \Omega \, \psi \;.
\end{equation}
The energy, $\Omega$, as seen by the rotating observer has been renamed in the mode functions to avoid confusing it with the previously defined $E$, which represented the Minkowskian observed energy.
These mode functions are orthogonal under the Dirac inner product \eqref{eq:diracprod} with the following normalisation
\begin{equation}
	\begin{aligned}
		&\langle \psi^\pm_{\Omega,\theta_p,p_z,s} , \psi^\pm_{\Omega',\theta_p',p_z',s'} \rangle \\
		&=\frac{1}{2\pi}\delta(\Omega - \Omega')\; \delta(p_z - p_z')\;\delta_{ss'} \; \sum_{l\le\xi_i,l\ge\xi_+}^\infty  e^{\mp i(l+\frac{1}{2})(\theta_p - \theta_p')} \;.
	\end{aligned}
\end{equation}
In the flat space scenario this normalisation can be written clearly as
\begin{equation}
		\langle \psi^\pm_{\Omega,\theta_p,p_z,s} , \psi^\pm_{\Omega',\theta_p',p_z',s'} \rangle = \delta(\Omega - \Omega')\; \delta(\theta_p - \theta_p') \; \delta(p_z - p_z')\;\delta_{ss'} \;,
\end{equation}
and behaves similarly in the rotating case. The missing $l$ modes in the summation have little consequence when performing Bogoliubov transformations or calculating expectation values.

We proceed to quantise the field using the standard canonical quantization procedure. Specifically, we construct the field operator with particles being defined with respect to the rotating energy, $\Omega$. 
\begin{equation}
	\begin{aligned}
	&\hat \Psi_R(x)\\
	&= \int_{\Omega \ge m}^\infty d \Omega \int_0^{2\pi} d\theta_p \int_{-p}^p dp_z \sum_{s=\pm 1} \left[\hat a_{j} \, \psi^+_j(x) + {\hat b_j}^\dagger \, \psi^-_j(x) \right] \;.
\end{aligned}
	\label{eq:field}
\end{equation}
The following standard equal-time anti-commutator relations have been imposed in promoting the field to an operator in \eqref{eq:field}
\begin{equation}
	\{ \hat \Psi_R(t,\textbf{x}), i \sqrt{-g} \, \hat \Psi_R^\dagger(t,\textbf{y}) \} = i \delta^3 (\textbf{x} - \textbf{y}) \;.
\end{equation}
These relations imply the following anti-commutators for the Fock space operators
\begin{equation}
	\begin{aligned}
	&\left\{ \hat a_j, {\hat a_{j'}}^\dagger \right\} = \\
	&\frac{1}{2\pi}\delta(\Omega - \Omega')\; \delta(p_z - p_z')\;\delta_{ss'} \; \sum_{l\le \xi_-,l\ge\xi_+}^\infty  e^{-i(l+\frac{1}{2})(\theta_p - \theta_p')} \\
	&\left\{ \hat b_j, {\hat b_{j'}}^\dagger \right\}= \\
	&\frac{1}{2\pi}\delta(\Omega - \Omega')\; \delta(p_z - p_z')\;\delta_{ss'} \; \sum_{l\le \xi_-,l\ge\xi_+}^\infty  e^{i(l+\frac{1}{2})(\theta_p - \theta_p')}\;,
	\end{aligned}
\end{equation}
with all others vanishing. 

With this quantisation, one can now calculate the radial current as was done previously in \eqref{eq:oldcurrent}. For illustrative purposes, we can normalise the radial current, $J^1$, by the current density, $J^0$, as an estimate for the radial velocity. Using non-rotating fields (i.e $\omega = 0$) one can write,
\begin{equation}
	\begin{aligned}
		v^r &= \frac{\langle 1_{\Omega,\theta,p_z,s} |\, \hat J^1\, | 1_{\Omega,\theta_p,p_z,s} \rangle}{\langle 1_{\Omega,\theta,p_z,s} |\, \hat J^0\, | 1_{\Omega,\theta_p,p_z,s} \rangle} \\
		& = \frac{\langle 1_{\Omega,\theta,p_z,s} |\,r\, \hat \Psi^\dagger \gamma^0 \gamma^1 \hat \Psi \, | 1_{\Omega,\theta_p,p_z,s} \rangle}{\langle 1_{\Omega,\theta,p_z,s} |\,r\, \hat \Psi^\dagger \hat \Psi \, | 1_{\Omega,\theta_p,p_z,s} \rangle} \\
		&= \frac{p_\perp}{\Omega} \cos(\theta - \theta_p) \;.
\end{aligned}
\end{equation}
This quantisation now gives a sensible result for the radial current and velocity, specifically the radial velocity expected from a free one-particle travelling with energy, $\Omega$, and parameterised momentum, $\theta_p$.

\section{Bogoliubov Coefficients}
\label{sec:Bogoliubov}
Given the field operator, \eqref{eq:field}, we can calculate the Bogoliubov coefficients which match the operators of the rotating frame's field, $\hat \Psi_R$, to that of a Minkowskian observers flat space field, $\hat \Psi_M$, via the relations
\begin{equation}
	\alpha_{j,j'} = \langle \psi_{R,j}^+, \psi_{M,j'}^+ \rangle, \quad \beta_{j,j'} = \langle \psi_{R,j}^-, \psi_{M,j'}^+ \rangle \;.
\end{equation}
The flat space field, $\hat \Psi_M$ can be trivially obtained from $\hat \Psi_R$ by setting $\omega =0 $. Recalling the set of quantum numbers, $j = \{\Omega, \theta_p, p_z,s\}$.
These coefficients can be calculated analytically (see Appendix \ref{app:bogoliubov}) and are found to be 
\begin{equation}
	\begin{aligned}
		\alpha_{j,j'} =& \frac{1}{2\pi} e^{i(\Omega-\Omega')t}\; \delta(p_z - p_z') \; \delta_{ss'}\\ &\times \sum_{l=-\infty}^\infty e^{i(l+\frac{1}{2})(\theta_p - \theta_p')} \, \delta(\hat E  -\Omega')\,  \\
		\beta_{j,j'} =& \frac{i}{2\pi} \frac{|\Omega'|}{\Omega'}\; \text{sgn}(s)  e^{-i(\Omega+\Omega')t} \; \delta(p_z + p_z') \; \delta_{ss'} \\ \times & \sum_{l=-\infty}^\infty e^{-i(l+\frac{1}{2})(\theta_p - \theta_p')}  e^{i l{\pi}} \; \delta(\hat E + \Omega')\; \;.
\end{aligned}
\label{eq:bogoliubov}
\end{equation}
Particle mixing/production occurs between the references frames when there is a non-zero $\beta$ Bogoliubov coefficient. With this quantisation, this occurs when $|\hat E| = |\Omega + \omega(l+\frac{1}{2})| \ge m$ and $\hat E = - \Omega'$.  
Recalling that the quantisation of the fields requires that $\Omega \ge m$, it is evident that the rotational effects (terms involving $\omega$) allow these conditions to be met and therefore give rise to particle production/mixing. With no rotation, this coefficient is zero and we will therefore see no inertial effects. 

These coefficients allow one to relate operators in the rotating frame, ($\hat a_j, \hat b_j$) with those of a stationary observer in flat space, denoted with capitals ($\hat A_j, \hat B_j$) via the following
\begin{equation}
        \begin{aligned}
                \hat{a}_{j} &=   \int_{-\infty}^{\infty} dj' \, \alpha_{j,j'} \hat{A}_{j'} + \beta^*_{j,j'} \hat{B}_{j'}^\dagger \\
		\hat{b}^\dagger_{j} &=  \int_{-\infty}^{\infty} dj' \, \beta_{j,j'} \hat{A}_{j'} + \alpha^*_{j,j'} \hat{B}_{j'}^\dagger  \;.
\end{aligned}
        \label{eq:bogrelations}
\end{equation}
Through these relations we can map rotating vacua to flat space vacua and vice versa. As the quanta in this quantisation are the generalisation to rotation of free flat space plane wave solutions, it is now possible to calculate expectation values in a rotating frame of single free one-particle states. An example is given in the following section. 

\section{Zitterbewegung in a Rotating Frame}
Zitterbewegung (``trembling motion'') of a free electron was theorised by Schr\"odinger in 1930 \cite{Schrodinger1930} and was erroneously due to calculations involving superpositions of positive and negative solutions of the Dirac equation. These superpositions can be realised between observers with differing quantisations of the Dirac field. An example of this is in an expanding spacetime \cite{Kobakhidze2015a}. Using this alternate quantisation, we determine if this ``trembling motion'' (found through calculating a one-particle state's velocity) can be observed by a rotating observer.    

In order to estimate the velocity of a particle as seen from a rotating observer one considers the conserved current operator 
\begin{equation}
	\hat J_R^\mu = \sqrt{-g}\, e_a^0 e_b^\mu,\hat \Psi_R^\dagger \gamma^a \gamma^b \hat \Psi_R \;.
\end{equation}
To determine the velocity expectation value of a Minkowskian 1-particle state as observed by a rotating observer, we calculate the following expectation value 
\begin{equation}
	\langle 1_{M,p} |:  \hat J_R^\mu :| 1_{M,p} \rangle  \;,
	\label{eq:currentexp}
\end{equation}
here we have used the notation $:\;:$ to represent normal-ordering with respect to the rotating operators. The $| 1_{M,p} \rangle$ represents a one-particle Minkowski particle with quantum numbers $p = \{\Omega_p,\theta_p,p_z,s\}$ in the Minkowskian Fock space denoted with subscript M's and vacuum $|0_M\rangle$. 

By relating the Minkowski states to rotating states using relations \eqref{eq:bogrelations}, we can write expectation values of the form \eqref{eq:currentexp} in the following way
\begin{equation}
	\begin{aligned}
	&\left< 1_p| :\hat{\Psi}_R^\dagger\, O^\mu \, \hat{\Psi}_R :| 1_p \right> =  \int \int dj \, dj' \\
	&{\psi^+_j}^\dagger O^\mu \, \psi^+_{j'} \;	\left[\alpha_{j,p}^* \alpha_{j',p} \, + \int_{-\infty}^{\infty} dk \, \beta_{j,k} \beta_{j',k}^* \delta(0) \right]	 & + \\
	&{\psi^+_j}^\dagger O^\mu \, \psi^-_{j'} \; \left[\alpha_{j,p}^* \beta_{j',p}	+ \int_{-\infty}^{\infty} dk \, \beta_{j,k} \alpha_{j',k}^* \delta(0) \right] &+	 \\
	&{\psi^-_j}^\dagger O^\mu \, \psi^+_{j'} \left[\beta_{j,p}^* \alpha_{j',p} + \int_{-\infty}^{\infty} dk \, \alpha_{j,k} \beta_{j',k}^* \delta(0) \right] &+	\\ 
	&{\psi^-_j}^\dagger O^\mu \, \psi^-_{j'} \left[\beta_{j,p}^* \beta_{j',p} - \int_{-\infty}^{\infty} dk \, \beta_{j,k}^* \beta_{j',k} \delta(0) \right]	\;.
	\label{eq:expandedform}
\end{aligned}
\end{equation}

The vector $O^\mu = \{r,r\gamma^0 \gamma^1, -r\omega + \gamma^0 \gamma^2, r\gamma^0 \gamma^3\}$ shows explicitly the individual operators required to calculate the four components of \eqref{eq:currentexp} for a rotating frame. All the divergent terms (terms containing $\delta (0)$) sum to zero, a result expected from normal ordering the operator in \eqref{eq:currentexp}. The remaining terms in equation \eqref{eq:expandedform} can be calculated analytically and represent contributions from positive and negative mode functions due to the non-zero $\beta$ coefficient. The calculations show that the rotation mixes up contributions from particles and anti-particles, however for this particular expectation value, the terms contribute in such a way that there is no net effect observed in the velocity. A sample calculation showing this for the $J^0$ current is given in Appendix\ref{app:velcal} and all other results are simply stated below. 
\begin{equation}
	\begin{aligned}
		J^0  &=E_p r \frac{1}{8 \pi^3} \xi_{\theta_p,p_\perp}(\theta,r) \\
	J^r &= p_\perp  r \frac{1}{8 \pi^3} \cos\left( \theta - \theta_p \right)\xi_{\theta_p,p_\perp} (\theta,r)\\
	J^\theta &=(-r\omega + p_\perp \sin(\theta - \theta_p ) ) \frac{1}{8 \pi^3}\xi_{\theta_p,p_\perp} (\theta,r)\\
	J^z &= r p_z \frac{1}{8 \pi^3} \sum_{l=-\infty}^{\infty} \xi_{\theta_p,p_\perp} (\theta,r) \;,
\end{aligned}
\end{equation}

where we have introduced the function
\begin{equation}
	\xi_{\theta_p,p_\perp}(\theta,r) = \sum_{l=-\infty}^{\infty} \sum_{l'=-\infty}^{\infty} e^{-i(l-l')(\theta - (\theta_p + \frac{\pi}{2}))} J_l(p_\perp r) J_{l'} (p_\perp r) \;.
\end{equation}
We can give an estimate to the velocity of a single-particle as seen by a rotating observer via 
\begin{equation}
	v^i = \frac{J^i}{J^0} \;.
\end{equation}
The estimates of the velocity in the rotating frame, gives 
\begin{equation}
	\begin{aligned}
		v^r &= \frac{p_\perp}{E_p} \cos(\theta -\theta_p) \\
		v^\theta &= -\omega + \frac{p_\perp}{rE_p} \sin(\theta -\theta_p) \\
		v^z &= \frac{p_z}{E_p} \;,
	\end{aligned}
\end{equation}
which amount to the standard classical expectation values for a rotating observer measuring a single particle state with parameterised momentum, $\theta_p$. The results show no explicit $\omega$ dependence in velocities other than the $\theta$ direction, which is expected classically. 

\section{Conclusion}  
Here we have presented a new quantisation for fermionic fields in cylindrical co-ordinates applicable to both flat and rotating metrics. Unlike common quantisations in these coordinates, our quanta have non-zero radial currents, which match to the expected radial component of a particle with a definite trajectory. This quantisation is formulated through the infinite sum of angular modes in previous calculations allowing the quanta to match the flat plane waves of cartesian quantisations.
We have shown the operators of the rotating field exhibit mixing through non-zero Bogoliubov coefficients to those of flat Minkowskian space. Despite this mixing, through explicit calculation, we have shown that there are no observable deviations in the velocity expectation of a particle when observed in a rotating frame. This occurs because the contributions in the expectation value from the $\beta$ Bogoliubov coefficients sum in such a way that any rotational dependence in the radial velocity expectation is cancelled. The same effect occurs in uniformly accelerating frames (see \cite{Kobakhidze2015a}). 

The Bogoliubov coefficients and field operators presented here are general, allowing one to calculate any expectation value in rotating or cylindrical systems using this new quantisation. 

\begin{acknowledgements} 
I would like to thank Archil Kobakhidze for many helpful discussions and advice in writing this paper. I would also like to thank Victor E. Ambrus for clarifying various gauge choices in these coordinate systems. This work was partially supported by the Australian Research Council. 
\end{acknowledgements}

\onecolumngrid
\appendix*
\section*{Appendix}

\subsection{Bogoliubov Calculation}
\label{app:bogoliubov}
This section demonstrates the calculation involved in obtaining the $\beta_{j,j'}$ Bogoliubov coefficient found in \eqref{eq:bogoliubov}.
This coefficient can be evaluated explicitly as

\begin{equation}
	\begin{aligned}
	\beta_{j,j'} &= \langle \psi_{R,j}^- | \psi_{M,j'}^+ \rangle\\
	&= \int_0^\infty \int_{0}^{2\pi} \int_{-\infty}^{\infty} \frac{r dr d\theta dz}{32\pi^3} \sum_{l\le\xi_-,l\ge\xi_+}\sum_{l'\le\xi_-,l'\ge\xi_+} \sqrt{\frac{|\hat E|}{\hat E}} \sqrt{\frac{| \Omega' |}{\Omega'}} e^{-i(\Omega + \Omega')t} e^{i(l+ \frac{1}{2})(\theta - (\theta_p + \frac{\pi}{2})) + (l' + \frac{1}{2})(\theta - (\theta_p' + \frac{\pi}{2}))} e^{i(p_z + p_z')} v(r)^\dagger u(r)
\end{aligned}
\label{eq:betacalc}
\end{equation}
Here $v(r)$ is the spinor component associated with the negative mode function, i.e $i\gamma^2\gamma^0{\gamma^0}^T u(r)^*$. 
Taking the $\theta$ and $z$ integrals, gives a Kronecker delta, $\delta_{l,-l'-1}$  allowing one to eliminate one of the $l$ sums and a standard delta function for the z-momentum. i.e
\begin{equation}
	\begin{aligned}
		\frac{1}{8 \pi} e^{-i(\Omega + \Omega')t} \sqrt{\frac{|\hat E|}{\hat E}} \sqrt{\frac{| \Omega' |}{\Omega'}} \sum_{l\le\xi_-,l\ge\xi_+}  e^{i(l+ \frac{1}{2})(\theta_p' - \theta_p)} \delta(p_z + p_z') \int_0^\infty r dr \; v(r)^\dagger u(r)
\end{aligned}
\end{equation}
Now the spinor components have the form
\begin{equation}
	v(r)^\dagger u(r) =
\begin{pmatrix} i \frac{|\hat E|}{\hat E} \frac{\sqrt{\hat E - ms_z} \sqrt{p_z + ms_z} }{\sqrt{ms_z}} J_{l+1} (p_\perp r) \\
		- \frac{\sqrt{\hat E + ms_z} \sqrt{p_z +ms_z}}{\sqrt{ms_z}} J_l (p_\perp r) \\
		i \frac{| \hat E|}{\hat E} \frac{\sqrt{m(\hat E - ms_z)}}{\sqrt{s_z(p_z + ms_z)}} J_{l+1}(p_\perp r) \\
		\frac{\sqrt{m(\hat E + ms_z)}}{\sqrt{s_z(p_z + ms_z}} J_l (p_\perp r)
	\end{pmatrix}^T 
	\begin{pmatrix}
		\frac{\sqrt{m(\Omega' + ms_z')}}{\sqrt{s_z'(p_z' + s_z')}} J_{-l -1} (p_\perp' r) \\
		-i \frac{|\Omega'|}{\Omega'} \frac{\sqrt{m(\Omega' - ms_z')}}{\sqrt{s_z'(p_z' + s_z')}} J_{-l} (p_\perp' r) \\
		\frac{\sqrt{\Omega' + ms_z'} \sqrt{p_z' + ms_z'}}{\sqrt{ms_z'}} J_{-l-1} (p_\perp' r )\\
		i \frac{|\Omega'|}{\Omega'}\frac{\sqrt{\Omega' - ms_z'} \sqrt{p_z' + ms_z'}}{\sqrt{m s_z'}} J_{-l} (p_\perp' r )
	\end{pmatrix}
\end{equation}
From these spinor relations, it is clear the $r$ integral will run over products of Bessel functions. There will be terms which contain,
\begin{equation}
	\begin{aligned}
&	\int_0^\infty r dr \;	J_{l}(p_\perp r) J_{-l}(p_\perp' r)  \\
=&	\int_0^\infty r dr \;	(-1)^{l} J_{l}(p_\perp r) J_{l}(p_\perp' r) \\
=& (-1)^{l} \frac{\delta(p_\perp - p_\perp')}{\sqrt{p_\perp p_\perp'}}
\end{aligned}
\end{equation}
In the second line we have used the Bessel function relation\cite{Gradshteyn2007}:
\begin{equation}
	J_{-n} (z) = (-1)^nJ_n (z) \;,
	\label{eq:besselinvert}
\end{equation} 
and have used the normalisation of Bessel functions to get from the second line to the third. We can re-write this delta function in terms of energies in the following way. First we recall that $p_\perp = \sqrt{(\Omega + \omega(l + \frac{1}{2}))^2 - m^2 -p_z^2}$ and $p_\perp' = \sqrt{\Omega'^2 - m^2 - p_z^2}$. Where we have imposed the delta function, $\delta(p_z + p_z')$, calculated earlier. 
Now, we have (for $x_i$ being zero's of $g(x)$): 
\begin{equation}
	\delta (g(x)) = \sum_i \frac{\delta(x - x_i)}{|g'(x_i)|}
\end{equation}
Hence
\begin{equation}
	\begin{aligned}		
&\frac{\delta(p_\perp - p_\perp')}{\sqrt{p_\perp p_\perp'}} \\
&= \frac{1}{\sqrt{p_\perp p_\perp'}}\frac{\sqrt{\Omega'^2 - m^2 -p^2}}{|\Omega'|} \left[ \delta(\hat E - \Omega') + \delta(\hat E + \Omega')  \right] \\
&= \frac{1}{|\Omega'|}\left( \delta(\hat E-\Omega') + \delta(\hat E+\Omega') \right)
	\end{aligned}
\end{equation}
The spin structure is such that the expectation value \eqref{eq:betacalc} is zero if $\hat E = \Omega'$ so the first delta function can be safely ignored. This ultimately leaves us with 

\begin{equation}
		\beta_{j,j'} = \frac{i}{2\pi} \frac{|\Omega'|}{\Omega'}\; \text{sgn}(s)  e^{-i(\Omega+\Omega')t} \; \delta(p_z + p_z') \; \delta_{ss'} \sum_{l\le\xi_-,l\ge\xi_+}^\infty e^{-i(l+\frac{1}{2})(\theta_p - \theta_p')}  e^{i l{\pi}} \; \delta(\hat E + \Omega')\; \;.
\end{equation}
For given energies, $\Omega$, the allowed $l$ sums grow. In a distributional sense, where these Bogoliubov coefficients are integrated over all allowed energies, the range of the $l$ sums become infinite, and so we write
\begin{equation}
		\beta_{j,j'} = \frac{i}{2\pi} \frac{|\Omega'|}{\Omega'}\; \text{sgn}(s)  e^{-i(\Omega+\Omega')t} \; \delta(p_z + p_z') \; \delta_{ss'} \sum_{l=-\infty}^\infty e^{-i(l+\frac{1}{2})(\theta_p - \theta_p')}  e^{i l{\pi}} \; \delta(\hat E + \Omega')\; \;.
\end{equation}

\subsection{Sample $J^0$ Calculation}
\label{app:velcal}
Here we present the explicit calculation of \eqref{eq:expandedform} for $J^0$, that is with $O = r$. Substituting the mode functions and Bogoliubov coefficients and performing the integrals, we achieve the following form
\begin{equation}
	\begin{aligned}
		J^0 =	& \frac{r}{16\pi^3} \Big[  \\
			&+\sum_{l=-\infty}^{\floor{\frac{E_p -m}{\omega} - \frac{1}{2}}} \sum_{l'=-\infty}^{\floor{\frac{E_p -m}{\omega} - \frac{1}{2}}} e^{-i(l-l')(\theta - (\theta_p + \frac{\pi}{2}))} \left[ (E_p + ms_z) J_l (p_\perp r)J_{l'} (p_\perp r)+ (E_p - ms_z) J_{l+1}(p_\perp r) J_{l'+1} (p_\perp r) \right] \\
			&+\sum_{l=-\infty}^{\floor{\frac{E_p -m}{\omega} - \frac{1}{2}}} \sum_{l'=-\infty}^{\floor{\frac{-E_p -m}{\omega} - \frac{1}{2}}} e^{-i(l+l'+1)(\theta - (\theta_p + \frac{\pi}{2}))} e^{il'\pi} \left[ (E_p - ms_z) J_{l'} (p_\perp r)J_{l+1} (p_\perp r)- (E_p + ms_z) J_{l'+1} (p_\perp r) J_{l}(p_\perp r) \right] \\
			&+\sum_{l=-\infty}^{\floor{\frac{-E_p -m}{\omega} - \frac{1}{2}}} \sum_{l'=-\infty}^{\floor{\frac{E_p -m}{\omega} - \frac{1}{2}}} e^{i(l+l'+1)(\theta - (\theta_p + \frac{\pi}{2}))} e^{-il \pi} \left[ (E_p - ms_z) J_l(p_\perp r) J_{l'+1}(p_\perp r) - (E_p + ms_z) J_{l'} (p_\perp r) J_{l+1} (p_\perp r) \right] \\
			&+\sum_{l=-\infty}^{\floor{\frac{-E_p -m}{\omega} - \frac{1}{2}}} \sum_{l'=-\infty}^{\floor{\frac{-E_p -m}{\omega}} - \frac{1}{2}} e^{i(l-l')(\theta - (\theta_p + \frac{3\pi}{2}))} \left[ (E_p + ms_z) J_{l+1} (p_\perp r)J_{l'+1} (p_\perp r)+ (E_p - ms_z) J_{l}(p_\perp) J_{l'} (p_\perp r) \right] \Big] \;.
	\end{aligned}
	\label{eqa:j0raw}
\end{equation}
The $\omega$ dependence originates from satisfying the energy delta functions in the Bogoliubov coefficients and become a restriction on the allowed $l$ terms.  The $\omega$ dependence also determines the contribution from each term in \eqref{eq:expandedform} through the limits in the $l$ summations. For example, in flat space with $\omega \rightarrow 0$ we get only a contribution from the first term with a summation going from $-\infty$ to $\infty$, implying $\beta_{j,j'} = 0$ and hence degeneracy between the rotating and flat space vacua.

In order to simplify \eqref{eqa:j0raw} we can use the Bessel function relation \eqref{eq:besselinvert} and make the substitution $l \rightarrow -l$, $l' \rightarrow -l'$ to re-write the last term as
\begin{equation}
	 \frac{r}{16\pi^3} \sum^{\infty}_{l=\ceil{\frac{E_p -m}{\omega} - \frac{1}{2}}} \sum^{\infty}_{l'=\ceil{\frac{E_p -m}{\omega} - \frac{1}{2}}} e^{-i(l-l')(\theta - (\theta_p + \frac{\pi}{2}))} \left[ (E_p + ms_z) J_l (p_\perp r)J_{l'} (p_\perp r)+ (E_p - ms_z) J_{l+1} (p_\perp r)J_{l'+1}(p_\perp r) \right] \;.
\end{equation}
Using a similar trick we can write the two middle terms as
\begin{equation}
	\begin{aligned}
		 \frac{r}{16\pi^3}\sum_{l=-\infty}^{\floor{\frac{E_p -m}{\omega} - \frac{1}{2}}} \sum^{\infty}_{l'=\ceil{\frac{E_p -m}{\omega} + \frac{1}{2}}} e^{-i(l-l'+1)(\theta - (\theta_p + \frac{\pi}{2}))} \left[ (E_p - ms_z) J_{l'} J_{l+1} + (E_p + ms_z) J_{l'-1} J_{l} \right] \\
		 \frac{r}{16\pi^3} \sum^{\infty}_{l=\ceil{\frac{E_p -m}{\omega} + \frac{1}{2}}} \sum_{l'=-\infty}^{\floor{\frac{E_p -m}{\omega} - \frac{1}{2}}} e^{i(-l+l'+1)(\theta - (\theta_p + \frac{\pi}{2}))} \left[ (E_p - ms_z) J_l J_{l'+1} + (E_p + ms_z) J_{l'} J_{l-1} \right] \;,
	\end{aligned}
\end{equation}
then by redefining $l$ and $l'$ as shifts by 1 to match the limits of the sums, we can write them as
\begin{equation}
	\begin{aligned}
		\frac{r}{16 \pi^3} \sum_{l=-\infty}^{\floor{\frac{E_p -m}{\omega} - \frac{1}{2}}} \sum^{\infty}_{l'=\ceil{\frac{E_p -m}{\omega} - \frac{1}{2}}} e^{-i(l-l')(\theta - (\theta_p + \frac{\pi}{2}))} \left[ (E_p - ms_z) J_{l'+1} (p_\perp r)J_{l+1} (p_\perp r)+ (E_p + ms_z) J_{l'} (p_\perp r)J_{l} (p_\perp r)\right] \\
		\frac{r}{16 \pi^3} \sum^{\infty}_{l=\ceil{\frac{E_p -m}{\omega} - \frac{1}{2}}} \sum_{l'=-\infty}^{\floor{\frac{E_p -m}{\omega} - \frac{1}{2}}} e^{-i(l-l')(\theta - (\theta_p + \frac{\pi}{2}))} \left[ (E_p - ms_z) J_{l+1} (p_\perp r)J_{l'+1} (p_\perp r)+ (E_p + ms_z) J_{l'} (p_\perp r)J_{l} (p_\perp r)\right] \;.
	\end{aligned}
\end{equation}
In total this allows us to write the four terms as
\begin{equation}
	\begin{aligned}
		\frac{r}{16 \pi^3}\bigg[ 	&+\sum_{l=-\infty}^{\floor{\frac{E_p -m}{\omega} - \frac{1}{2}}} \sum_{l'=-\infty}^{\floor{\frac{E_p -m}{\omega} - \frac{1}{2}}} e^{-i(l-l')(\theta - (\theta_p + \frac{\pi}{2}))} \left[ (E_p + ms_z) J_l (p_\perp r)J_{l'} (p_\perp r)+ (E_p - ms_z) J_{l+1} (p_\perp r)J_{l'+1} (p_\perp r)\right] \\
			&+\sum^{\infty}_{l=\ceil{\frac{E_p -m}{\omega} - \frac{1}{2}}} \sum^{\infty}_{l'=\ceil{\frac{E_p -m}{\omega} - \frac{1}{2}}} e^{-i(l-l')(\theta - (\theta_p + \frac{\pi}{2}))} \left[ (E_p + ms_z) J_l (p_\perp r)J_{l'} (p_\perp r)+ (E_p - ms_z) J_{l+1} (p_\perp r)J_{l'+1} (p_\perp r)\right] \\
		&+\sum_{l=-\infty}^{\floor{\frac{E_p -m}{\omega} - \frac{1}{2}}} \sum^{\infty}_{l'=\ceil{\frac{E_p -m}{\omega} - \frac{1}{2}}} e^{-i(l-l')(\theta - (\theta_p + \frac{\pi}{2}))} \left[ (E_p - ms_z) J_{l'+1} (p_\perp r)J_{l+1} (p_\perp r)+ (E_p + ms_z) J_{l'} (p_\perp r)J_{l} (p_\perp r)\right] \\
		&+\sum^{\infty}_{l=\ceil{\frac{E_p -m}{\omega} - \frac{1}{2}}} \sum_{l'=-\infty}^{\floor{\frac{E_p -m}{\omega} - \frac{1}{2}}} e^{-i(l-l')(\theta - (\theta_p + \frac{\pi}{2}))} \left[ (E_p - ms_z) J_{l+1} (p_\perp r)J_{l'+1} (p_\perp r)+ (E_p + ms_z) J_{l'} (p_\perp r)J_{l} (p_\perp r)\right] \bigg],
	\end{aligned}
\end{equation}
which we can write simply as the double summation
\begin{equation}
	\frac{E_p r}{8 \pi^3} \sum_{l=-\infty}^{\infty} \sum_{l'=-\infty}^{\infty} e^{-i(l-l')(\theta - (\theta_p + \frac{\pi}{2}))} J_l (p_\perp r)J_{l'}(p_\perp r) \;.
\end{equation}
Through this calculation we can see the total contributions from each term in \eqref{eq:expandedform}, regardless of $\omega$, will sum up to a double infinite sum just as if we had originally taken $\omega \rightarrow 0$. This is the reason we see no observable effect between rotating and inertial frames.

\twocolumngrid

\end{document}